\documentclass[runningheads]{llncs}
\usepackage{graphicx}
\usepackage{hyperref}

\usepackage{amsmath}
\usepackage{amsfonts}
\usepackage{multirow}
\usepackage{cancel}
\usepackage{subcaption}
\usepackage{marvosym}
\usepackage{booktabs}
\begin{document}

\title{DeReStainer: H\&E to IHC Pathological Image Translation via Decoupled Staining Channels}
%
\titlerunning{DeReStainer}
%
\authorrunning{Wei et al.}
\author{Linda Wei ${ }^1$, Shengyi Hua ${ }^1$, Shaoting Zhang ${ }^{2}\textsuperscript{(\Letter)}$, Xiaofan Zhang ${ }^{1}$}

\institute{${ }^1$ Shanghai Jiao Tong University\\
${ }^2$ Shanghai Artificial Intelligence Laboratory\\
}
\maketitle              
\begin{abstract}
    Breast cancer is a highly fatal disease among cancers in women, and early detection is crucial for treatment. HER2 status, a valuable diagnostic marker based on Immunohistochemistry (IHC) staining, is instrumental in determining breast cancer status. The high cost of IHC staining and the ubiquity of Hematoxylin and Eosin (H\&E) staining make the conversion from H\&E to IHC staining essential. In this article, we propose a destain-restain framework for converting H\&E staining to IHC staining, leveraging the characteristic that H\&E staining and IHC staining of the same tissue sections share the Hematoxylin channel. We further design loss functions specifically for Hematoxylin and Diaminobenzidin (DAB) channels to generate IHC images exploiting insights from separated staining channels. Beyond the benchmark metrics on BCI contest, we have developed semantic information metrics for the HER2 level. The experimental results demonstrated that our method outperforms previous open-sourced methods in terms of image intrinsic property and semantic information.
\keywords{H\&E-to-IHC Stain Translation \and Gerative Adversarial Network  \and  Staining Separation}
\end{abstract}
\section{Introduction}
Breast cancer (BC) is a highly prevalent and deadly cancer among females \cite{breastcancer}. The mortality rate of early-stage breast cancer is significantly lower than that of the advanced stages \cite{riihimaki2012death}. Consequently, early detection of BC is essential for preventing its progression and substantially decreasing the risk of mortality in patients.

\begin{figure}[t]
    \begin{subfigure}[b]{0.55\textwidth}
        \includegraphics[width=\textwidth]{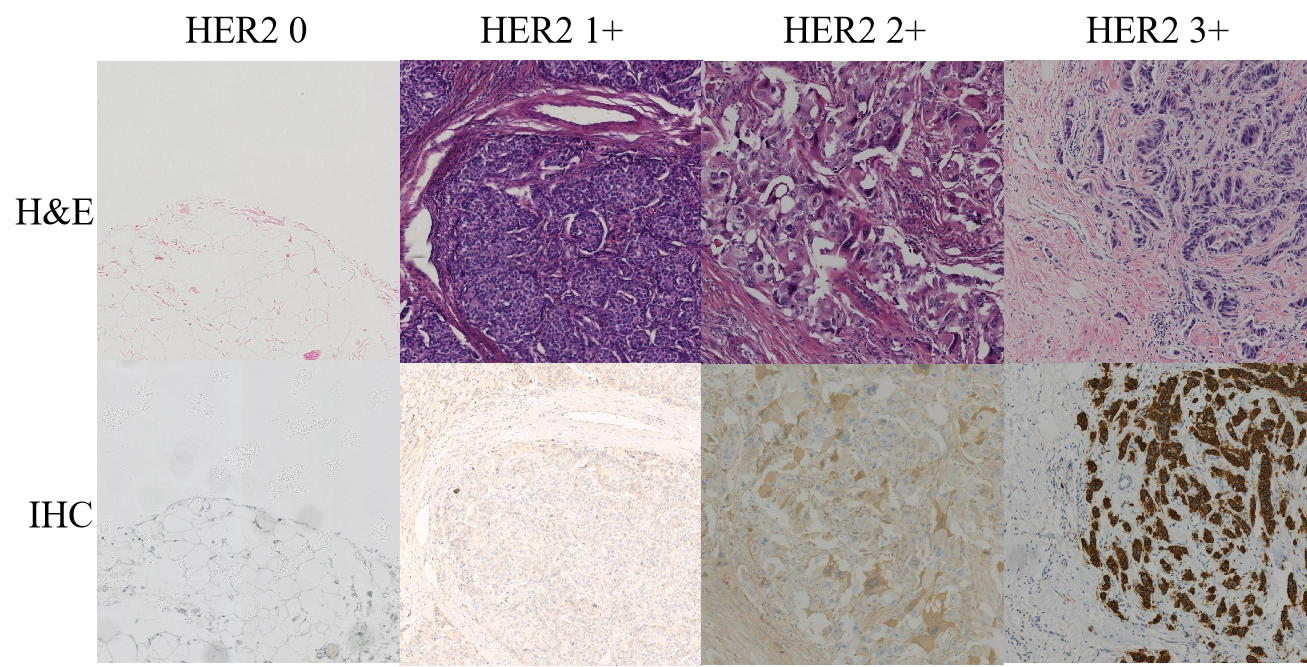}
        \caption{}
        \label{fig:heihc}
    \end{subfigure}
    \hfill
    \begin{subfigure}[b]{0.435\textwidth}
        \includegraphics[width=\textwidth]{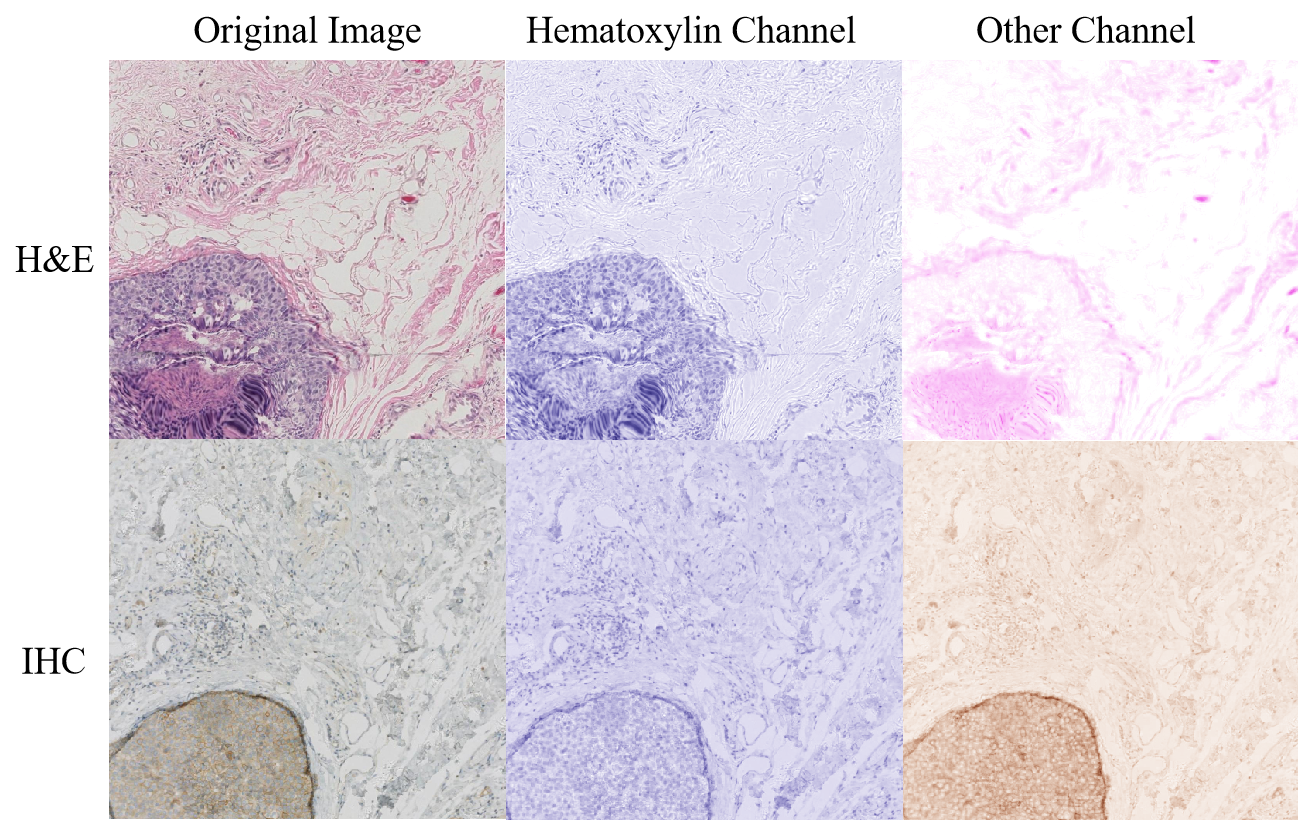}
        \caption{}
        \label{fig:separation}
    \end{subfigure}
    \caption{(a): Examples of H\&E and IHC pairs of different HER2 levels.(b): Stain Separation for H\&E and IHC image. The first column is the original images. The second column is the shared H channel. The third column is the other channel, E channel for H\&E, and DAB channel for IHC}
    \label{fig:Fig1}
\end{figure}

Immunohistochemistry (IHC) staining plays a pivotal role in pathology, particularly in the identification of tumor cells \cite{ihc}. 
IHC staining uses antibodies applied to tissues that target and bind to antigens in tumor cells. Once bound, these antibodies are visualized with various labeling techniques, enabling the detection and localization of specific cellular proteins. A key application is identifying HER2 (Human Epidermal Growth Factor Receptor 2), a protein that promotes cancer cell growth in BC. IHC allows for the quantification of HER2 levels, categorized as 0, 1+, 2+, or 3+ \cite{yamauchi2008her2} (see \autoref{fig:heihc}), which is vital for selecting an effective treatment strategy.
Despite the valuable diagnostic insights provided by IHC, the staining process is labor-intensive and time-consuming \cite{anglade2020can}. Therefore, developing a framework for the automatic generation of IHC image admits its significance.

Hematoxylin and Eosin (H\&E) staining, the gold standard for examining tissue structure and morphology stands as the most prevalent staining technique in pathology. Often, the choice to proceed with IHC staining is guided by the insights gained from H\&E-stained specimens. Moreover, numerous studies \cite{basavanhally2013multi,jenkins2007pathology} suggest that H\&E images hold crucial information relevant to IHC outcomes, supporting the theoretical possibility of predicting IHC expressions from H\&E data in medical research. Thus, the development of an algorithm capable of converting H\&E to IHC stained images emerges as a logical step towards automated IHC image generation.
The generation of an IHC image from its corresponding H\&E slice is an image-to-image translation (I2IT) task. There are many benchmark frameworks for this task in deep learning. 
Pix2Pix\cite{pix2pix} employed a cGAN for effective image translation in paired datasets. Building upon this foundation. Pix2PixHD \cite{pix2pixhd} enhanced the resolution of generated images with a multi-scale architecture. In medical image analysis, specialized models have been developed to address the H\&E to IHC translation, emphasizing the complexity and clinical relevance of pathological images. Liu et al. launched the Breast Cancer Immunohistochemical (BCI) competition and introduced the Pyramid Pix2Pix model which imposed stricter constrain on structural similarity for the staining translation task in H\&E images \cite{liu2022bci}. Li et al. applied PatchNCE loss to mitigate dataset alignment issues in BCI dataset \cite{li2023adaptive}. Despite achieving commendable results on structural metrics such as the Structural Similarity Index (SSIM) and the Peak Signal-to-Noise Ratio (PSNR) which are pivotal for assessing images' intrinsic properties, we observed that the IHC images produced by previous methods sometimes do not fully reflect the HER2 level information of the ground truth IHC images. This can lead to a less precise representation of the HER2 level in the pathology tissue, which may affect clinical significance.
Therefore, building on the work of predecessors, we further leverage the shared characteristics of H\&E and IHC staining to achieve effective translation from one to the other, with the aim that the generated IHC images more accurately reflect semantic information, specifically the HER2 level. 
From the perspective of staining principles, H\&E stained images contain two channels: the Hematoxylin (H) channel and the Eosin (E) channel. IHC-stained images use Diaminobenzidine (DAB) as a chromogenic substrate to mark the position of antigens and are counterstained with Hematoxylin to highlight the location of cell nuclei. Therefore, it can be viewed as the combination of H and DAB channels,
the staining separation result for H\&E and its corresponding IHC image is shown in \autoref{fig:separation}. 
Therefore, we believe that \textbf{in H\&E and IHC stained sections of the same tissue, the H channel is shared, with difference arising solely from the E and DAB channels.} 
Our framework decouples the H channel from an H\&E image and aligns it with the H channel of the corresponding IHC image in a high-dimensional feature space. In addition, it supervises the DAB channel of the generated IHC image, which further improves the semantic information of the generated IHC. In addition, we find that the original metrics for the BCI contest, the SSIM and PSNR values emphasized the evaluation of images' intrinsic properties, which cannot fully reflect the semantic information of the IHC image, especially the HER2 level. Therefore, we introduce metrics that emphasize the evaluation if the semantic information of the generated IHC image to assess our methodology. 

\noindent Our contributions in this work can be summarized as follows:
\begin{itemize}
\item We propose DeReStainer, a framework that leverages the shared H channel between H\&E and IHC stained images, achieving H\&E to IHC conversion through a destain-restain process.

\item Utilizing staining separation, we develop a loss function based on the DAB channel of IHC images to generate IHC images with accurate HER2 levels.

\item Beyond the SSIM and PSNR, we introduce metrics aiming to evaluate the semantic information of generated IHC images in our experiments, providing sufficient evidence to demonstrate the clinical significance of our method.

\end{itemize}
 \section{Methods} 
\begin{figure}
    \centering
    \includegraphics[width=\textwidth]{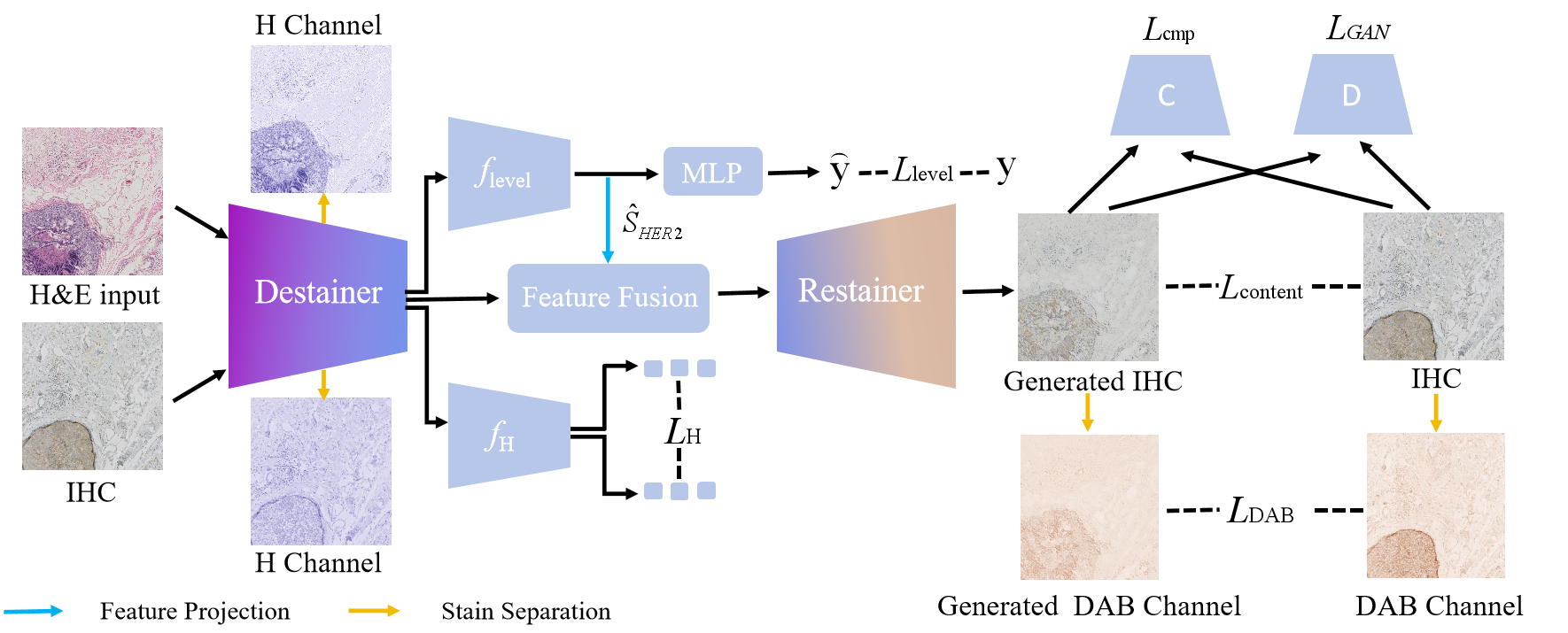}
    \caption{The overall architecture of our framework. DeStainer performs stain separation and aligns the H channel of H\&E with that of IHC images. The Feature Fusion Module integrates the output of DeStainer with the classification information of the HER2 level. Restainer is used to generate the corresponding IHC images. In the framework, y is the HER2-level label of the input IHC. $f$ represents a simple network, C and D are comparator and discriminator respectively. }
    \label{model}
\end{figure}

As is shown in \autoref{model}, our proposed DeReStainer performs a de-re stain process to generate IHC from $H\&E$images. The DeStainer aligns the H channel that is shared between H\&E and IHC images, while the ReStainer is responsible for applying DAB staining to generate the corresponding IHC images. 
Without further notice, we denote the H\&E image, generated IHC image, and the ground truth IHC image as $I_{H\&E}, \hat{I}_{IHC}, I_{IHC}$ respectively. 

\subsection{DeStainer}
The DeStainer contains a staining separation module and an encoder, which aim to embed the H channel of both H\&E and IHC images into a high-dimensional feature space and align them with each other.

\textbf{Staining Separation.} To get the H channel from the corresponding H\&E and IHC images, these images are first sent to the staining separation module. The staining separation module uses a staining separation algorithm based on color space transform \cite{ruifrok2001quantification} to separate input pathology images into the H channel and the other channels (E channel for H\&E and DAB channel for IHC image). Specifically, it first maps a tissue pathology image in the RGB space to the HED (Hematoxylin, Eosin, DAB) space. A projection matrix is then applied to map the image in HED space to a certain channel by setting the other two channels of this image to zero. After that, the resulting image is transformed back to RGB space to get the separated channel result. 
 For instance, the procedure to get the H channel of an H\&E image is shown in the following formula, where
 $T(\cdot):\mathbb{R}^{n\times n\times 3} \rightarrow \mathbb{R}^{n\times n\times 3}$ is the transform from RGB to HED space, $P^H \in \text{M}_3(\mathbb{R})$ represents projection matrix to H channel. The $I_{H\&E}^H$ here is transformed back to RGB space thus it is still 3-channel image. 
$$
\begin{aligned}
& I_{H\&E}^{HED}= \left[v_H, v_E, v_{DAB}\right]=T(\left[v_R, v_G, v_B\right])=T(I_{H\&E}^{R G B}) \in \mathbb{R}^{n\times n\times 3}\\ 
& I_{H\&E}^H=\left[v_R^{\prime} , v_G^{\prime}, v_B^{\prime}\right]=T^{-1}(\left[v_H, 0,0\right]) =T^{-1}(\left[v_H , v_E, v_{DAB}\right] \cdot P^H) \in \mathbb{R}^{n\times n\times 3}
\end{aligned}
$$
\textbf{Encoder.} $I_{H\&E}^H$ and $I_{IHC}^H$ from the staining separation module are sent into the encoder of DeStainer to get their the high-level representations. After that, these high-level representations are sent into $f_{H}$ to be aligned with each other. We designed $L_{{H}}$ to constrain this process, which is a cosine similarity loss defined as:
$$
L_{{H}} = 1-\frac{\langle f_{H}(\text{DeStainer}(I_{H\&E}^H), f_{H}(\text{DeStainer}(I_{IHC}^H))\rangle}{|f_{H}(\text{DeStainer}(I_{H\&E}^H)|\cdot|f_{H}(\text{DeStainer}(I_{IHC}^H))|}
$$

\subsection{Feature Fusion Module}
The feature fusion module aims to combine the feature of $I_{H\&E}^H$ and the HER2 level information to help generate IHC images with precise HER2 level. 

Before being sent into the feature fusion module, the high-level feature of $I_{H\&E}^H$ is directed to $f_{H}$ to get classification information for HER2 level, denoted as $\hat{S}_{HER2}$. Then, $\hat{S}_{HER2}$ will sent jointly with the output of DeStainer to the feature fusion model to get the input of ReStainer. It is modified from typical ResNet blocks and the convolutional layer in it is modified to modulation convolutional block in StyleGAN2 \cite{stylegan}, which performs weight demodulation operation.
The demodulation operation of modulation convolutional block can be formulated as
\begin{equation*}
   \text{ModConv}(x,\hat{S}_{HER2}) = \text{Conv}\left( x, \frac{W \times (\hat{S}_{HER2} + 1)}{\sqrt{\sum(W \times (\hat{S}_{HER2} + 1))^2 + \epsilon}} \right) + b , 
\end{equation*}
where $\texttt{Conv}$ is 2-d convolution, $W$ is the original weight of the convolutional layer, $b$ is an optional bias, and $\epsilon$ is a small constant.
Beyond that, SimAM \cite{simam}, a parameter-free attention layer is appended on the end of each block. 

\subsection{ReStainer}
The ReStainer is a decoder that is symmetric to the encoder of the DeStainer. The goal is to apply DAB staining on the output of the feature fusion module to generate the corresponding IHC images. We employ a staining separation module identical to that in the DeStainer to obtain $\hat{I}_{IHC}^{DAB}$ and $I_{IHC}^{DAB}$. To further improve staining quality, we have formulated 
$ L_{{DAB}}$, which is defined as:
$$
L_{{DAB}} = \|\hat{I}_{IHC}^{DAB}-I_{IHC}^{DAB}\|_1
$$
the DAB channel of the generated and ground truth IHC image are obtained from the same staining separation operation in DeStainer, but letting $v_H,v_E=0$ instead. Beyond the ReStainer, we incorporate a comparator and a discriminator into our framework. The comparator aims to bring the generated IHC images closer to real IHC images on the feature level, further strengthening the control of visual perception. The discriminator applies generative adversarial learning to make the generated IHC images more realistic. 
\subsection{Loss Functions}
\noindent \textbf{Overall Loss.}
The overall loss function is formulated as follows:
$$
L= \lambda_{stain}L_{stain}+\lambda_{content}L_{content}+\lambda_{level}L_{level}+\lambda_{GAN}L_{GAN}, 
$$
which contains $L_{stain}$ and auxiliary loss including $L_{content}, L_{level},$ and $L_{GAN}$.

\noindent \textbf{Stain Loss. } The stain loss is the most significant loss in our framework, which is the weighted sum of $ L_{{H}}$ and $ L_{{DAB}}$, whose purpose and formulation have been discussed in the former section respectively. 
The stain loss is formulated as:
$$
L_{stain} = \lambda_{H}L_{H}+\lambda_{DAB}L_{DAB}.
$$
\textbf{Auxiliary Losses. }Besides the stain loss, we also incorporates auxiliary losses to facilitate the generation of IHC images.
The content loss comprises structural similarity loss $L_{ssim}$ \cite{ssim}, mean absolute error loss $L_{mae}$, and comparator loss $L_{cmp}$. $L_{ssim}$ and $L_{mae}$ are calculated directly between $\hat{I}_{IHC}$ and $I_{IHC}$, while $L_{cmp}$ represents the cosine similarity loss between features of $\hat{I}_{IHC}$ and $I_{IHC}$, as extracted by comparator C.
The content loss can be formulated as
$
L_{content} = \lambda_{ssim}L_{ssim} + \lambda_{mae}L_{mae} + \lambda_{cmp}L_{cmp}
$.
The HER2 level loss $L_{level}$ is a focal loss\cite{lin2017focal} applied to $y$ to obtain more precise HER2-level information for the feature fusion module.
For the GAN loss $L_{GAN}$, the multi-scale patch GAN loss \cite{pix2pixhd} is employed as the loss function, which is the mean value of the patch GAN loss \cite{pix2pix} in resolution of 512 and 256.

 \section{Experiment}
\subsection{Dataset and Implementation Details}
\textbf{Dataset} The BCI contest dataset \cite{liu2022bci} is used in our experiment. This publicly available dataset consists of 3896 H\&E-HER2 and corresponding IHC pairs in the training set and 977 pairs for the testing. All the patches in this dataset are of the size $1024 \times 1024$.

\noindent \textbf{Implementation Detail of the framework.} Our framework was implemented on an NVIDIA GTX 3090 GPU and was trained on $512 \times 512$ patches. The generated IHC images were resized to $1024 \times 1024$ for evaluation. We used AdamW to train our framework and the initial learning rate is $1e-5$. For the staining separation module, $T$ was a composite of a logarithmic function and a non-singular linear transformation. The encoder in DeStainer contained 3 down-sampling blocks and $f_H,f_{level}$ contained 2 down-sampling blocks. Each downsampling block contained a convolutional layer, Batch Normalization, and LeakyReLU, reducing the resolution by half. he feature fusion module contained 9 modified ResNet blocks. The comparator was a pre-trained ResNet 50 on IHC images, which was trained using a classification task on IHC levels and achieved an accuracy of 94\%. 
The weight constants in the overall loss were $\lambda_{stain}=2, \lambda_{content}=13, \lambda_{level}=5, \lambda_{GAN}=1$. Within each kind of loss, the weight constants are $\lambda_{H}=1$, $\lambda_{DAB}=1$, $\lambda_{cmp}=2$, $\lambda_{mae}=10$, $\lambda_{ssim}=1$,  $\lambda_{level}=5$.
\subsection{Results}
\noindent \textbf{Evaluation Metrics.}
In the experiment, we evaluate the generated IHC images from two distinct perspectives. Firstly, the assessment focuses on the intrinsic properties of the images, employing benchmarks established in the BCI challenge, specifically the SSIM and the PSNR. These metrics compare the generated images to their corresponding ground truths, providing a quantifiable measure of image quality and structural integrity.
However, achieving excellent results solely on metrics for image intrinsic property is insufficient for clinical purposes. Moreover,  the physical structures of the ground truth and H\&E input in the dataset do not completely align, using SSIM and PSNR metrics may not directly reflect the quality of the generated results. In fact, The IHC images in the BCI dataset are characterized not only by their visual quality but also by their semantic content, particularly the HER2 level, which is crucial for clinical interpretations. Consequently, we introduced metrics for semantic evaluation. 
We employ a pre-trained ResNet-50 classifier on IHC images different from what the comparator is trained on to encode test set images into a feature library, i.e a list contains all the embedded feature of ground truth IHC images. For each generated image, we also use it to form a vector and assess the top-k HER2 level accuracy with $k \in \{1, 3, 5\}$. 
In addition to the top-k accuracy, we also apply the k-Nearest Neighbor(kNN) method to measure the HER2 level classification accuracy of $\hat{I}_{IHC}$. 
\begin{figure}[t]
    \centering
    \begin{subfigure}[b]{0.49\textwidth}
        \includegraphics[width=\textwidth]{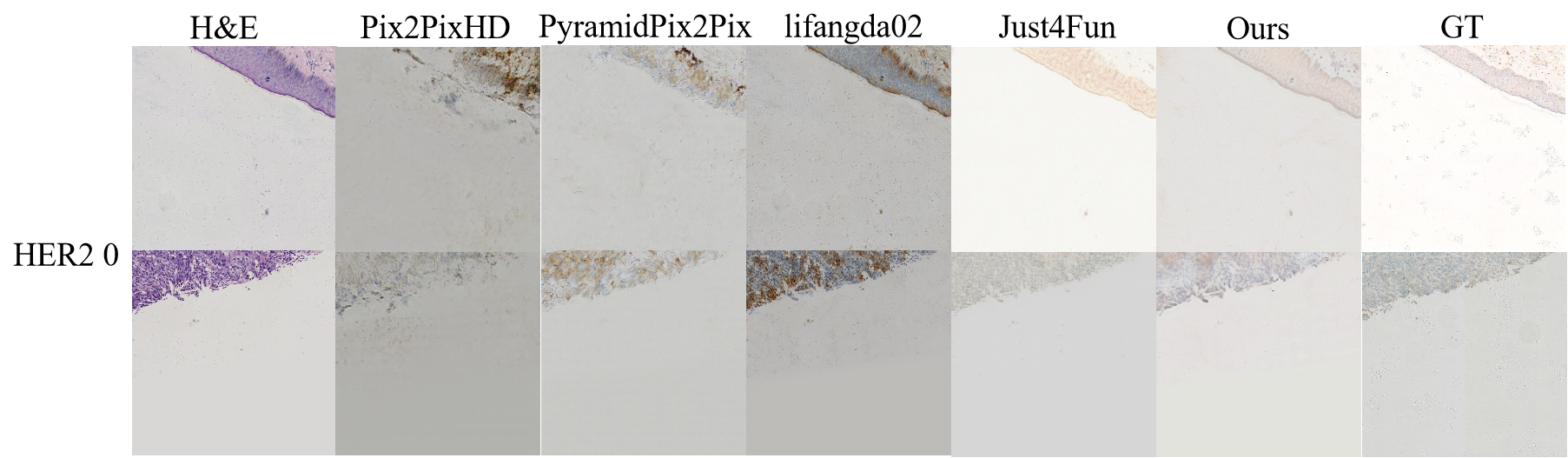}
        \label{fig:sub1}
    \end{subfigure}
    \hfill
    \begin{subfigure}[b]{0.49\textwidth}
        \includegraphics[width=\textwidth]{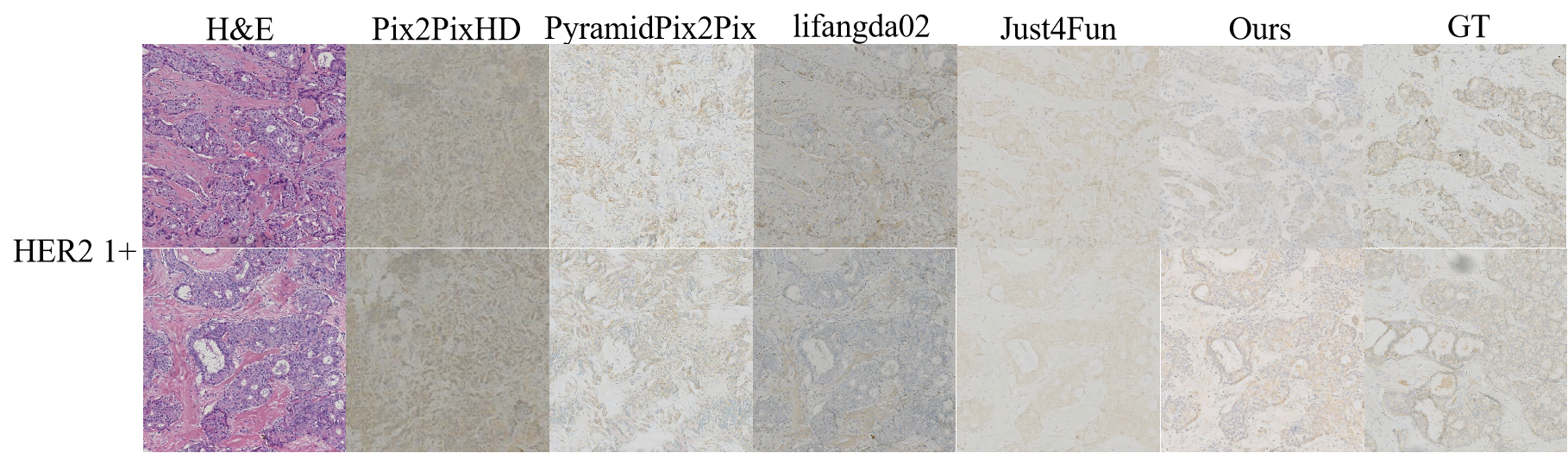}
        \label{fig:sub2}
    \end{subfigure}
    \newline
    \begin{subfigure}[b]{0.49\textwidth}
        \includegraphics[width=\textwidth]{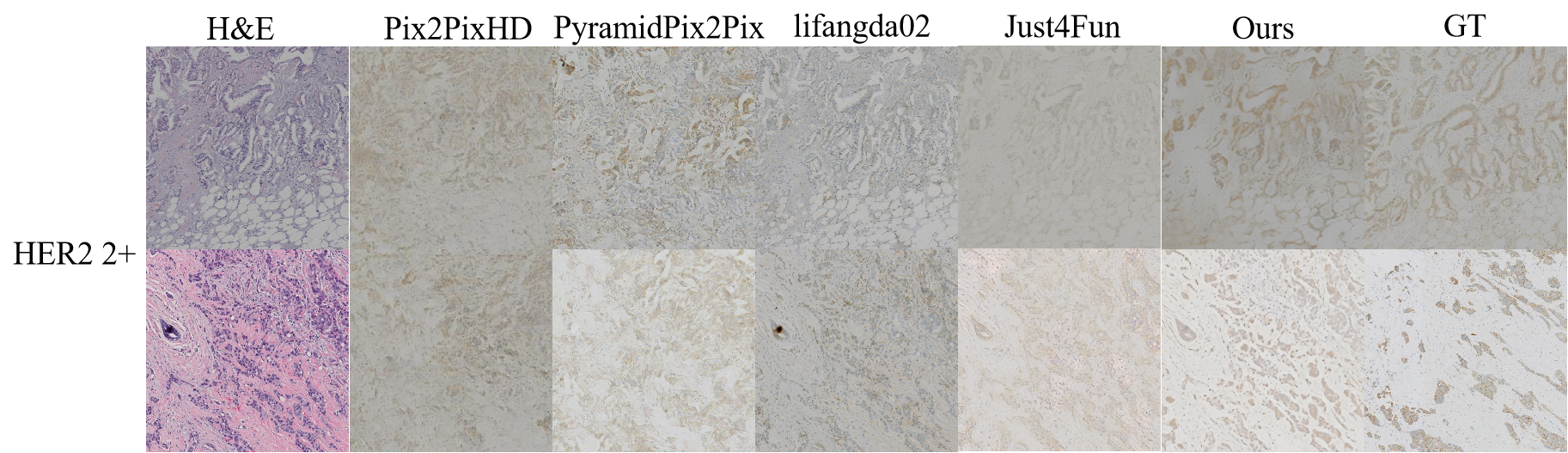}
        \label{fig:sub3}
    \end{subfigure}
    \hfill
    \begin{subfigure}[b]{0.49\textwidth}
        \includegraphics[width=\textwidth]{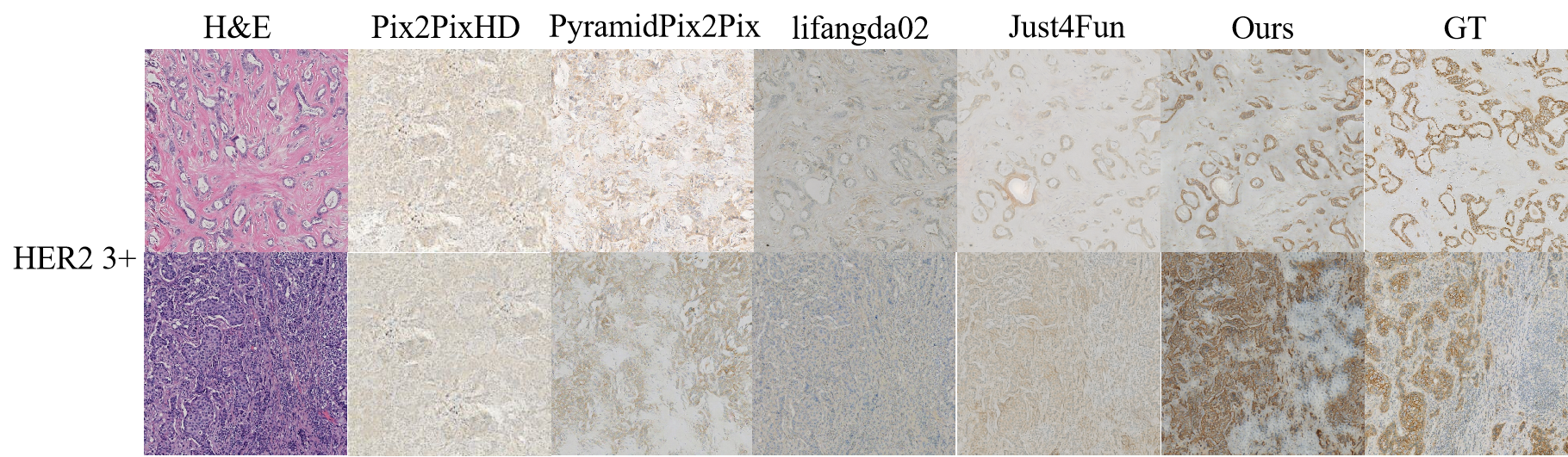}
        \label{fig:sub4}
    \end{subfigure}
    \caption{Visualization of generated IHC with all HER2 levels of different methods  }
    \label{fig:visualization}
\end{figure}

\noindent \textbf{Comparison Result.} 
\autoref{fig:visualization} shows the visual comparison results of our method against other open-sourced methods. Since the H\&E and corresponding IHC images in the BCI dataset are adjacent slices but cannot achieve perfect physical alignment, significant structural variances are observed between some of the generated images and the ground truth.
Hence, an effective result ought to exhibit structural similarity to the original H\&E image while mirroring the texture and color of the ground truth IHC image. The visual outcomes demonstrate that our method exceeds prior approaches in achieving structural clarity and color fidelity in IHC staining, especially in IHC with higher HER2 level (2+ and 3+).
\begin{table}[htbp]
\centering
\caption{Methods Comparison Result and Ablation Study}
\label{tab:comprehensive}

\begin{subtable}{\textwidth}
\centering
\caption{Comparison of Different Methods}
\label{tab:comparison}
\begin{tabular}{@{}lcccccc@{}}
\toprule
\multirow{2}{*}{Methods}       & \multirow{2}{*}{SSIM}  & \multirow{2}{*}{PSNR}   & \multicolumn{3}{c}{Top-k acc} & kNN \\ 
\cmidrule(lr){4-6} \cmidrule(lr){7-7}
              &       &        & k=1   & k=3   & k=5   &      k=10            \\ 
\midrule
Pix2PixHD \cite{bcireview}     & 0.471 & 19.634 & 0.276 & 0.480 & 0.563 & 0.278                 \\
PyramidPix2Pix \cite{liu2022bci} & 0.477 & 21.161 & 0.335 & 0.505 & 0.613 & 0.330             \\
Just4Fun \cite{bcireview}  & 0.558 & \textbf{22.929} & 0.483 & 0.670 & 0.779 & 0.490             \\
lifangda02 \cite{li2023adaptive}            & 0.503 & 17.865 & 0.368 & 0.543 & 0.624 & 0.353            \\
Ours          & \textbf{0.564} & 20.647 & \textbf{0.563} & \textbf{0.726} & \textbf{0.796} & \textbf{0.600}              \\
\bottomrule
\end{tabular}
\end{subtable}

\begin{subtable}{\textwidth}
\centering
\caption{Ablation Study of Staining Loss}
\label{tab:ablation}
\begin{tabular}{@{}lcccccc@{}}
\toprule
\multirow{2}{*}{Methods} & \multirow{2}{*}{SSIM} & \multirow{2}{*}{PSNR} & \multicolumn{3}{c}{Top-k acc} & kNN  \\ 
\cmidrule(lr){4-6} \cmidrule(lr){7-7}
        &      &      & k=1  & k=3  & k=5  & k=10 \\ 
\midrule
w/o $L_{stain}$ & 0.559 & \textbf{20.862} & 0.481 & 0.696 & 0.787 & 0.513 \\
w/o $L_{DAB}$    & 0.555 & 20.141 & 0.554 & 0.714 & 0.788 & 0.562 \\
Full $L_{stain}$       & \textbf{0.564} & 20.647 & \textbf{0.563} & \textbf{0.726} & \textbf{0.796} & \textbf{0.600} \\
\bottomrule
\end{tabular}
\end{subtable}

\end{table}
\autoref{tab:comparison} presents the results of our experiments. We compare our method with benchmark I2IT methods on natural images \cite{pix2pixhd} and the top-ranking open-source methods in the BCI leaderboard \cite{bcireview,li2023adaptive}, our method achieves the highest score in SSIM and also performs well in PSNR. Our method does not achieve SOTA in PSNR maybe because we used images of size 512 to train the framework instead of 1024. Apart from the SSIM and PSNR metrics, we also evaluate our model on semantic information metrics that reflect the HER2-level information of the generated IHC images. We calculate the top-k accuracy when $k=1,3,5$. In the kNN analysis, we employ $k=10$ to optimize majority voting accuracy during cross-validation. The results show that we have made significant improvements in semantic information metrics compared to previous methods. This underscores the potential of our method to generate IHC images with accurate HER2 levels, offering clinically more relevant results compared to previous methods.
\subsection{Ablation Study}

\autoref{tab:ablation} showcases our ablation study. In this work, we have primarily employed staining separation techniques to design the Hematoxylin loss and DAB loss, simulating the process of de-staining and re-staining. From \autoref{tab:ablation}, it is evident that these two losses play a role in the preservation of semantic information. In terms of PSNR, our full model experienced only a slight decrease, which has virtually no impact on the quality of the generated images.
However, upon integrating these two losses, the SSIM, top-k, as well as kNN accuracy of HER2 level in the generated IHC images, see a considerable increase. Moreover, the introduction of the DAB loss further improves the classification precision. 
The visualization result of the ablation study is shown in the supplementary material.

 \section{Conclusion}
 In this study, we utilize staining separation techniques based on the principle of destaining and restaining to develop a stain transfer framework and introduce new metrics to evaluate the semantic information of generated IHC images. Our framework not only demonstrates outstanding performance in metrics for image intrinsic properties such as SSIM and PSNR, but also achieves the best performance in top-k accuracy and kNN accuracy metrics based on HER2 level.

\bibliographystyle{splncs04}
 \bibliography{parts/ref}

\end{document}


\title{Supplementary Materials}

\author{Linda Wei ${ }^1$, Shengyi Hua ${ }^1$, Shaoting Zhang ${ }^{2}$, Xiaofan Zhang ${ }^{1}\textsuperscript{(\Letter)}$}

\institute{${ }^1$ Shanghai Jiao Tong University\\
${ }^2$ Shanghai Artificial Intelligence Laboratory\\
}
\maketitle

\section{Supporting images for Ablation Study}
\begin{figure}
    \centering
    \includegraphics[width=\textwidth]{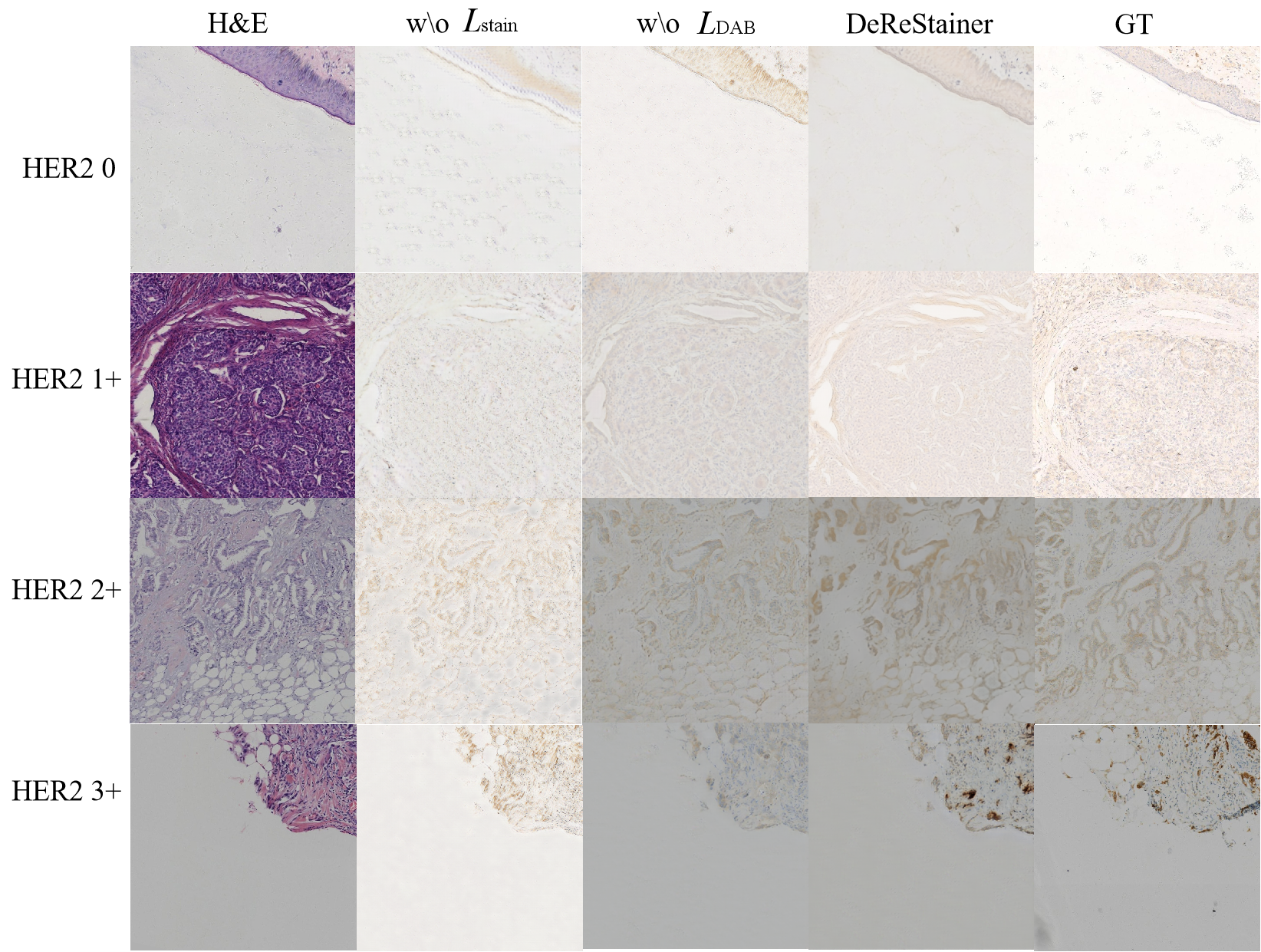}
    \caption{Supporting images for Ablation Study}
    \label{fig:enter-label}
\end{figure}